# Halcyon - A Pathology Imaging and Feature analysis and Management system

Authors: Erich Bremer, Tammy DiPrima, Joseph Balsamo, Jonas Almeida, Rajarsi Gupta, and Joel Saltz

Erich Bremer (corresponding author)
erich.bremer@stonybrook.edu
Stony Brook University
Biomedical Informatics

Tammy DiPrima
tammy.diprima@stonybrook.edu
Stony Brook University
Biomedical Informatics

Joseph Balsamo
joseph.balsamo@stonybrook.edu
Stony Brook University
Biomedical Informatics

Jonas Almeida
jonas.dealmeida@nih.gov
National Cancer Institute
Division of Cancer Epidemiology & Genetics

Rajarsi Gupta
Rajarsi.Gupta@stonybrookmedicine.edu
Stony Brook University
Biomedical Informatics

Joel Saltz
Joel.Saltz@stonybrookmedicine.edu
Stony Brook University
Biomedical Informatics

**Abstract**
Halcyon is a new pathology imaging analysis and feature management system based on W3C linked-data open standards and is designed to scale to support the needs for the voluminous production of features from deep-learning feature pipelines. Halcyon can support multiple users with a web-based UX with access to all user data over a standards-based web API allowing for integration with other processes and software systems. Identity management and data security is also provided.

**Introduction**
Deep-learning/AI pipelines have extracted and continue to extract tremendous amounts of morphological features from pathology whole slide images creating the daunting task of curating, storing, retrieving, and displaying the extracted features in clinically meaningful ways. The bulk of this data can be represented as polygons. Searching and filtering of spatial features creates some interesting and problematic challenges. Halcyon takes a novel approach representing polygon datasets as a series of Hilbert curve segments and stores this data in Apache Arrow files stored in self-describing Research Object Crate (RO-Crate) zip files {Soiland-Reyes, 2022 #37}. This data and the corresponding whole slide image (WSI) data are retrieved using the International Image Interoperability Framework (IIIF) and overlaid together in multiple user-specified colored feature layers using interlinked browser-based viewers. Halcyon provides researchers and clinicians a way to view complex feature result sets in an integrated fashion allowing for the simultaneous exploration of multiple classes of features and probability maps.

High-speed whole slide image (WSI) scanners are producing large volumes of high resolution images for Pathology applications {Wang, 2012 #8}. Any particular WSI can contain 500,000 to two million individual nuclei of various types with larger structures to be classified and annotated. Comparing and visualizing derived nuclear features for specific images and collections of images requiring a scalable instructure to store, manage, and access this information. Our earlier system QuIP stores derived feature data with spatial information and annotations represented as polygons in GeoJSON {Force, 2016 #27} using MongoDB as the NoSQL backend database {DICOM Standards Committee, 2021 #28}. Specific polygons in this system are retrieved using simple range requests configured to retrieve all polygons whose vertices fall within a particular $x_{min}/x_{max}, y_{min}/y_{max}$ range. The scale dependency of this approach comes with a cost in storage (indexing) further complicated by multiple regions of interest (ROI) annotations being themselves delimited by polygons at different scales. Furthermore, since all WSI's polygons are stored in the same database, indexing needed to cope with annotation grows rapidly with the volume of data. Keep in mind, ROI's are the areas we qualify, using polygons, as something important to the task at hand. This could include detailed segmentations such as the perimeters of nuclei or it could include coarse aggregate areas such as a tile-based approach using collections of square polygons to simplify processing.

Image viewing technology for WSI across technological solutions may rely on different serializations but generally follow similar memory-intensive approaches requiring RAM memory >32GB in many cases. Whole slide visualization systems (such as QuIP) support interactive WSI visualization by using image tiling servers to traverse the image pyramids used to represent WSI image files {Adelson, 1984 #29}. These pyramids provide the original images at various scales. If the whole image is required, the lowest resolution can be retrieved from the file without having to download and decode the rest of the file [{Bremer, 2020 #1}]. As the user zooms into a whole slide image, the tiling server loads tiles from successively higher resolution image pyramid layers. As the user zooms in, the tiles retrieved correspond to smaller and smaller spatial regions. This creates the "infinite zooming" experience supported by computational libraries used for layered navigation of maps and other images composed of large numbers of tiles at different scales, such as OpenSeadragon [{, #17}]. Formats such as GeoJSON {Force, 2016 #27} that represent polygons typically employ representations of vertices and edges. Polygon representations of this kind can increase data bandwidth requirements needed to support interactive zooming creating a laggy user interaction, as well as, a less visible but equally expensive computational inefficiencies in the integration of AI applications {Le, 2020 #12}.

The scale dependent tile pyramid approach to image representation is the ubiquitous model used in biomedical informatics applications {Goode, 2013 #19}. We propose an alternative feature image representation approach based on Hilbert Curves {Hilbert, 1935 #40}{Peano, 1990 #39}. Hilbert curves provide an interesting alternative by using its space-filling property to map from two dimensional space to one dimensional space with desirable locality preserving properties {Moon, 2001 #13}. This fractal matrix convolution approach creates a single coordinate system that maps all pixels of an image in a manner where the relative length of each position is scale independent {Liang, 2008 #11}. For example, a position at ⅓ of the length of a Hilbert curve will point to the same location of the image regardless of the resolution of the polygons.

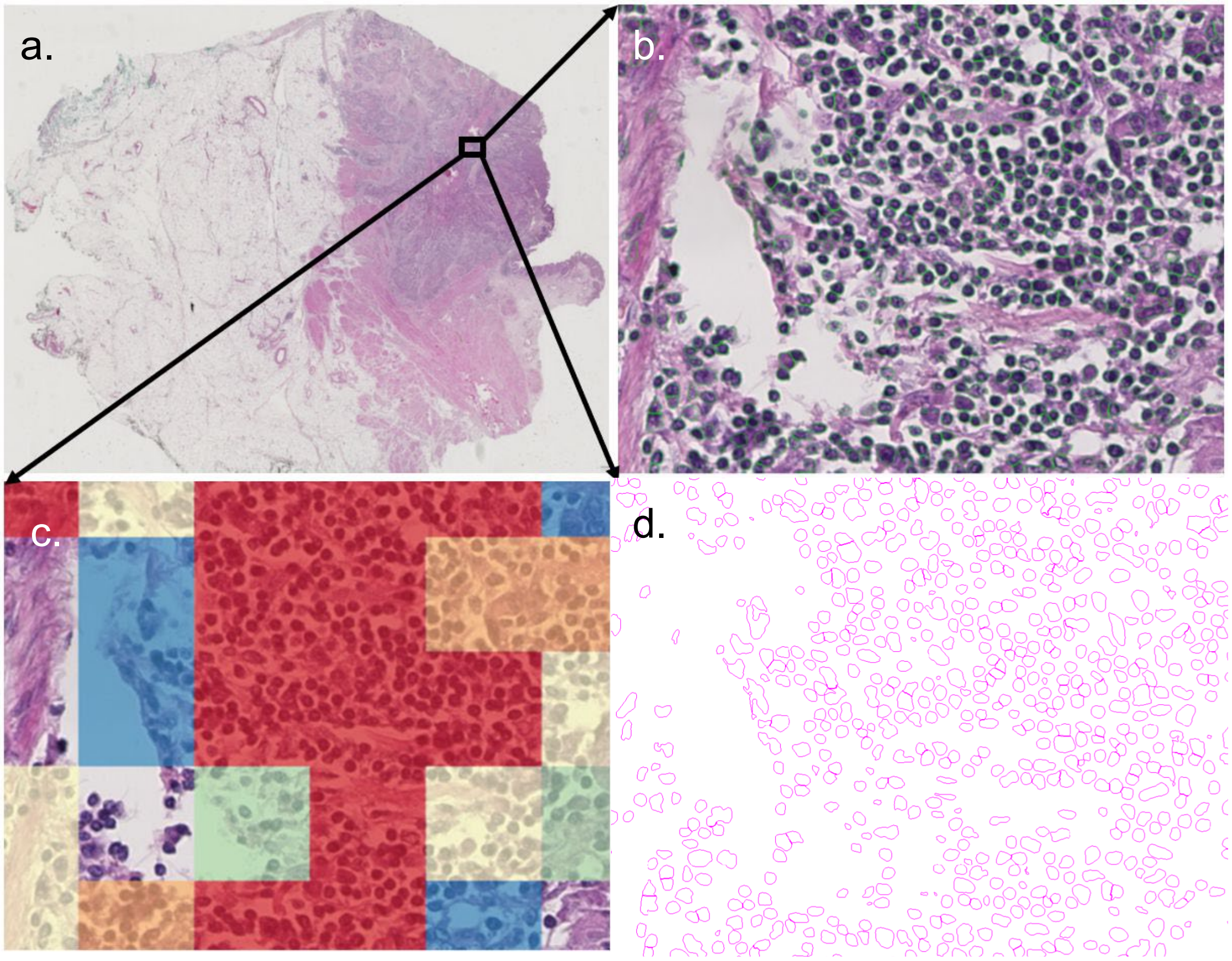

**Figure 1 - ROI in a Cancer whole slide image (a). Nuclear material ROI overlaid on the same image (b), nuclear material ROIs only (d), and a heatmap representation for tumor infiltrating lymphocytes (c).**

**Technical Background**

Polygons are represented in various ways, usually, as sequentially connected points along the perimeter of the polygon the last point either implicitly connected to the first point in the list or the first point is explicitly listed at the end of the list. The region inside this closed region would represent the region of interest.  This is represented in various ways including but not limited the following textually notations:

JSON (JavaScript Object Notation)
{“coordinates”: [[[1,1],[1,4],[3,5],[5,3],[4,1],[1,1]]], “Type”: “Polygon”}

Scalable Vector Graphics (SVG)
<svg><polygon points=”1,1 1,4 3,5 5,3 4,1 1,1” style="fill:lime;stroke:purple;stroke-width:1"/></svg>

Well-known text representation (WKT)
POLYGON (1 1, 1 4, 3 5, 5 3, 4 1, 1 1)

An alternate method for ROI representation would be as a bitmap array with “zero” to represent background and a “one” to indicate a piece of the ROI.

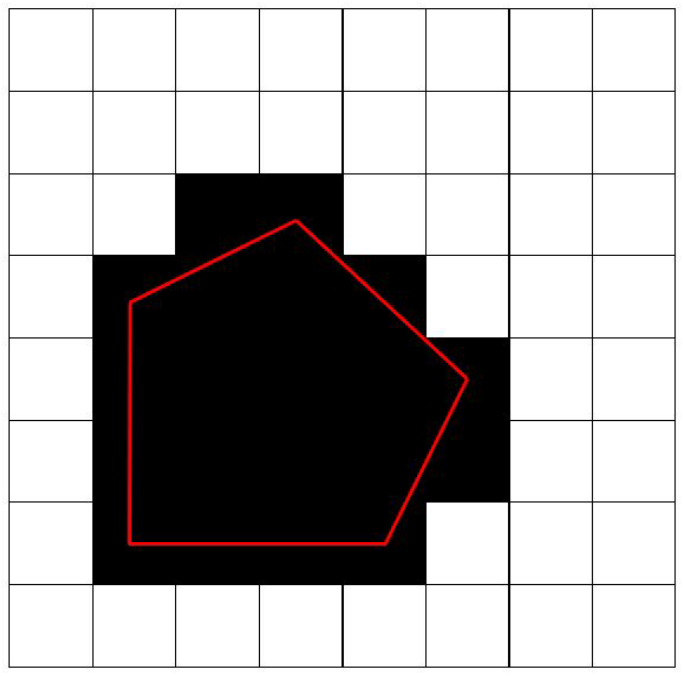

**Figure 3 - bitmap representation of a polygon**

At first glance, this would seem like a potential loss of resolution, however, due to requirements of various machine vision algorithms, it is often the bitmap that is the original data product with the textual representation being derived from the bitmap. Many software libraries exist to convert between these two representations as both polygon representations are often both employed.

As with most data, polygons are loaded into a database where spatial queries are done to subset that data as required. Not all databases employ advanced spatial indexing methods but all will support simple range queries for spatial object selection. For a particular rectangular search area S->($S_{minx}$ , $S_{miny}$ , $S_{maxx}$ , $S_{maxy}$) with $S_{min}$ and $S_{max}$ being the upper left and lower right corners of the rectangular ROI (Figure 1), a SQL query would look as follows:

select polygon.id, polygon.x, polygon.y where
((polygon.x BETWEEN $S_{minx}$ AND $S_{maxx}$ ) and (polygon.y BETWEEN $S_{miny}$ AND $S_{maxy}$ ))

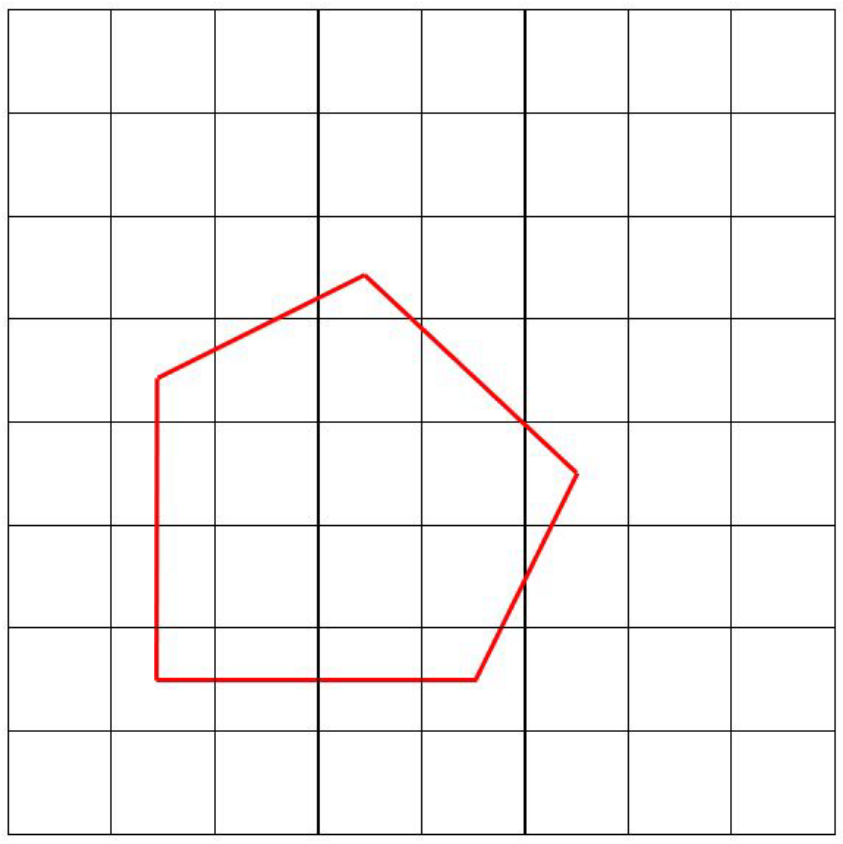

**Figure 2 - simple irregular polygon**

This approach works but performance erodes as billions of spatial objects in the database are approached depending on server CPU and IO horsepower. There are many spatial indexing algorithms such as geohash, hhcode, z-order curve, quadtree, octree, UB-tree, R-tree, R+ tree, R* tree, x-tree, kd-tree, m-tree, and bsp trees. Each has its advantages and disadvantages including algorithm coding complexity. For this paper, the Hilbert Curve was selected not only as a spatial indexing method but for actually representing the polygon itself rather than a cartesian perimeter form or as a binary mask.

A Hilbert curve (Figure 4) is a continuous fractal space-filling curve that was first described in 1891 by German mathematician David Hilbert {Hilbert, 1935 #40}. In any rectangular grid or array, a Hilbert curve will visit each and every cell once and only once without the curve crossing over on itself. Hilbert curves give a mapping between 2D and 1D space that preserves a degree of locality{Moon, 2001 #13}.

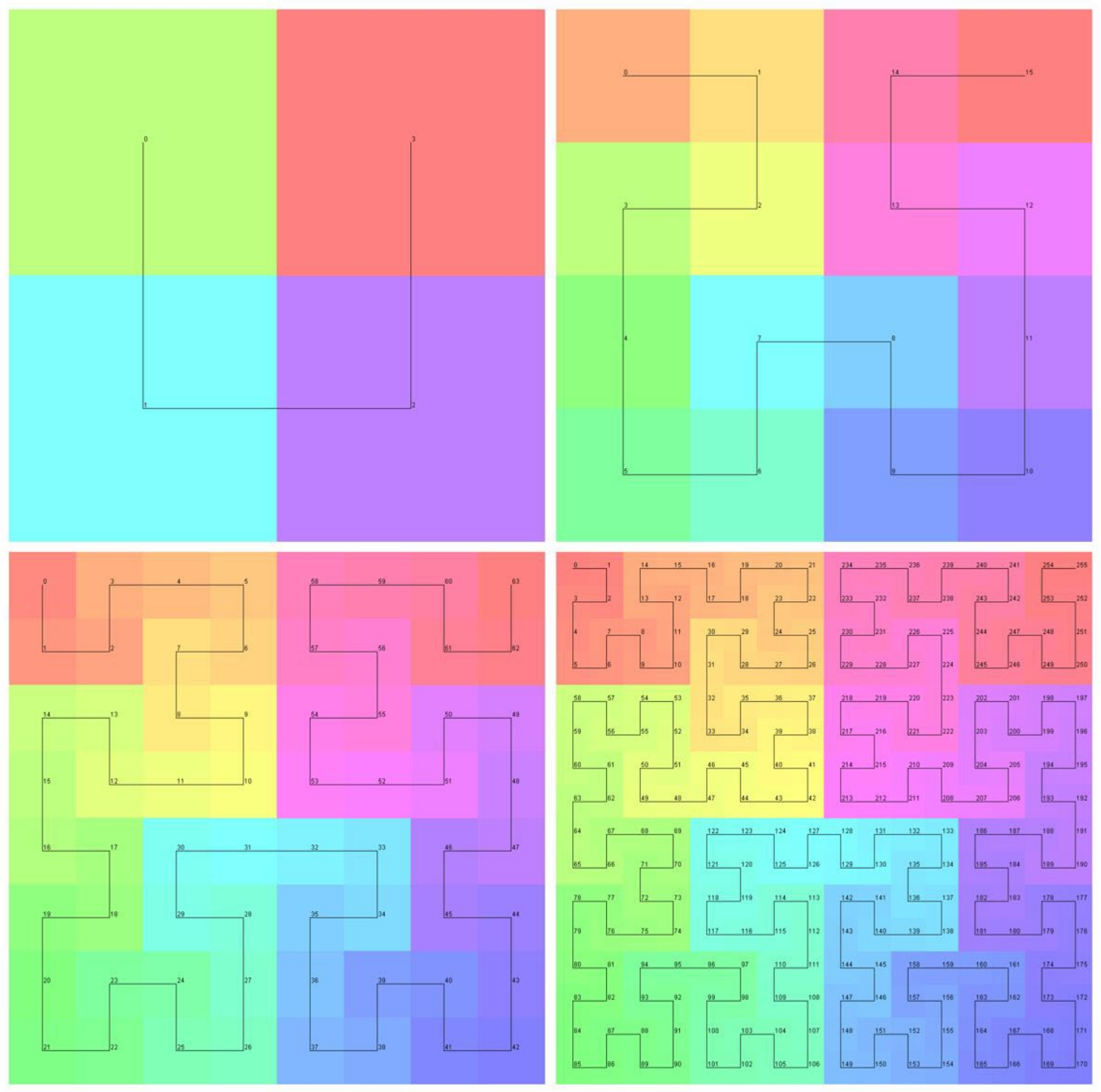

**Figure 4 - First, 2nd, 3rd,  and 4th order Hilbert Curves**

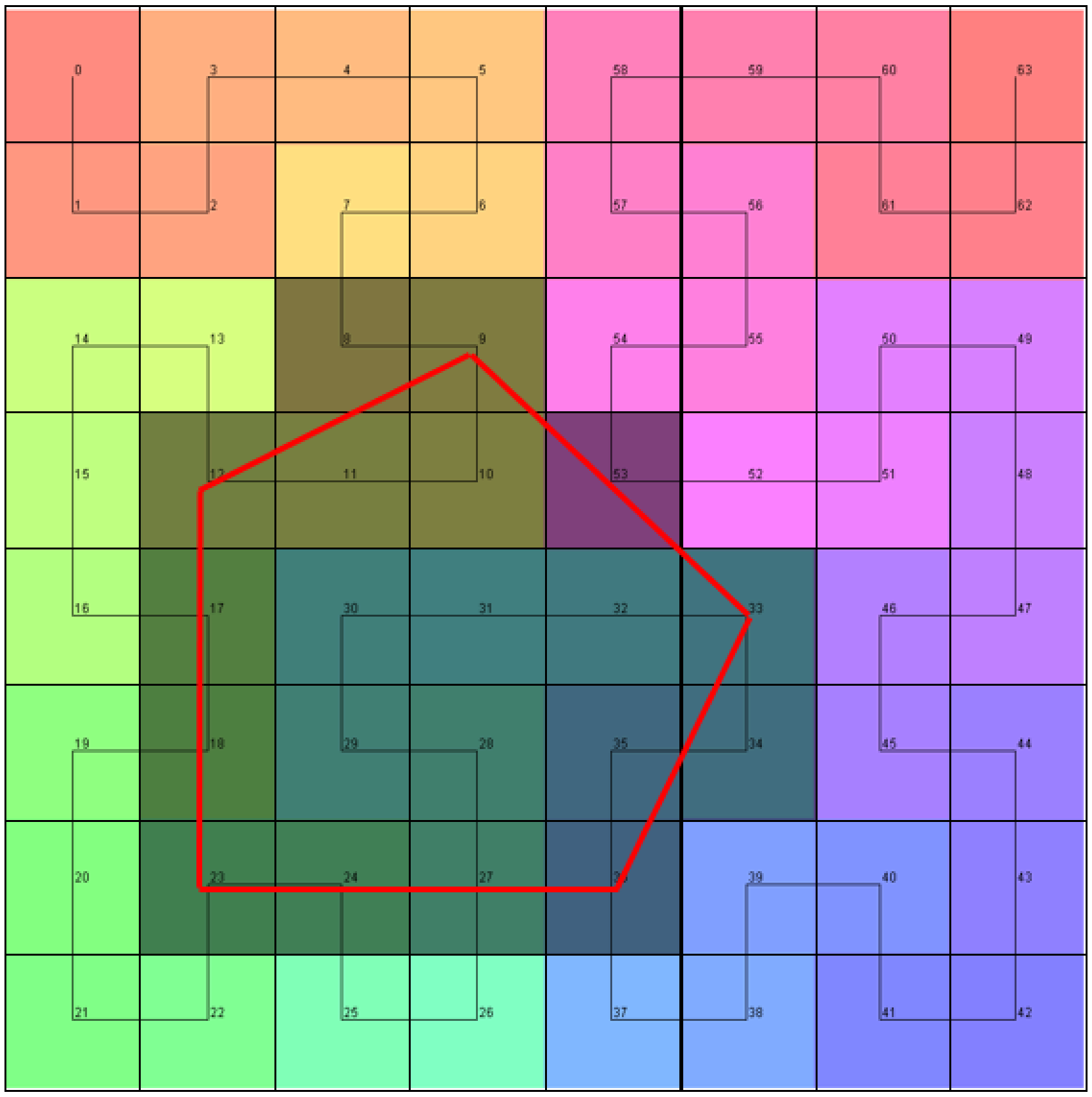

**Figure 5 - Polygon overlaid on a Hilbert Curve**

Hilbert Polygon Representation in JSON for the above polygon (Figure 5)

{"name": "Polygon 1",
"type": "Nuclear Material",
"Ranges":[[8,12],[17,18],[23,24],[27,36],[53,53]]}

In RDF Turtle:

[ a <http://snomed.info/id/4421005>;
:ranges
[ :low 8; :high 12], [ :low 17; :high 18], [ :low 23; :high 24], [ :low 27; :high 36], [ :low 53; :high 53]]

This representation of the polygon can be converted back by simply looking at the perimeter of the enclosed region. This is derived from the boundary algorithm for finding ranges from Moten, "Given an n-dimensional search region the exact hilbert curve ranges that cover the search region can be determined just by looking at the hilbert curve values on the perimeter (boundary) of the region."{Moten, #33} For our purposes, we store our polygons solely in Hilbert form, search with Hilbert ranges, and only when the cartesian representation is needed, we use this algorithm to convert our Hilbert representations back to their cartesian versions.

Casting the two-dimensional points onto the one dimensional Hilbert curve has the critical advantage that it can be layered into any database. Not all databases support spatial indexing but all support one-dimensional numeric indexes. Using this strategy, we can map each Hilbert Polygon range into a flat table that contains the beginning and end for each range and the ID of that range which can be mapped back to another table which contains the ID of the Hilbert Polygon which that range is a part of. The table is sorted on the field for the value of the start and end Hilbert ranges.

Take our original rectangular search area, S->($S_{minx}$ , $S_{miny}$ , $S_{maxx}$ , $S_{maxy}$). This search area itself is a polygon and can also be converted to a Hilbert Polygon $S_{hp}$ ->($S_{1[b1,e1]}$, $S_{2[b2,e2]}$, $S_{3[b3,e3]}$, $S_{4[b4,e4]}$, $S_{5[b5,e5]}$, ...) containing one or more being/end ranges. To obtain the list of polygons in S hp , we simply take the union of the following:

select polygon.id, polygon.x, polygon.y where
(polygon.hilbertvalue BETWEEN $S_{b1}$ AND $S_{e1}$ ) UNION

select polygon.id, polygon.x, polygon.y where
(polygon.hilbertvalue BETWEEN $S_{b2}$ AND $S_{e2}$ ) UNION ...

select polygon.id, polygon.x, polygon.y where
(polygon.hilbertvalue BETWEEN $S_{bn}$ AND $S_{en}$ ) UNION …

The Hilbert Curve polygon representation method was applied to an WSI image from The Cancer Genome Atlas (TCGA) collection for the use case of nuclear segmentation. The irregular polygons extracted from the segmentation progress creates the most challenge to this methodology. Here below is a comparison of the two methods:

Standard connected perimeter point polygons method
The dimensions of the TCGA image were: 135,168 x 105,472 pixels
# of polygons extracted for nuclear segmentation = 1,547,170
# of points required to represent standard polygons = 54,600,980
35.3 (X,Y) points per polygon

Hilbert representation method
# of Hilbert ranges = 36,478,264
23.6 Hilbert Ranges per polygon

**Halcyon General Architecture**
Halcyon is designed to collect, organize, and provide access to whole slide images (WSI) and visualize their derived feature sets simultaneously overlaid on their original source WSI.

**Figure 6 - Halcyon Overall Design Schema**

**Data Representation**
Halcyon's data storage is based on W3C Linked Data Resource Description Framework (RDF) Model {, 2014 #35} {Barros, 2016 #25} {García-Closas, 2023 #41}. RDF provides a graph-based model with multiple name spaces supported through the use of URIs.  W3C recommended or well-established ontologies are used, when available, as descriptors for the various data elements.

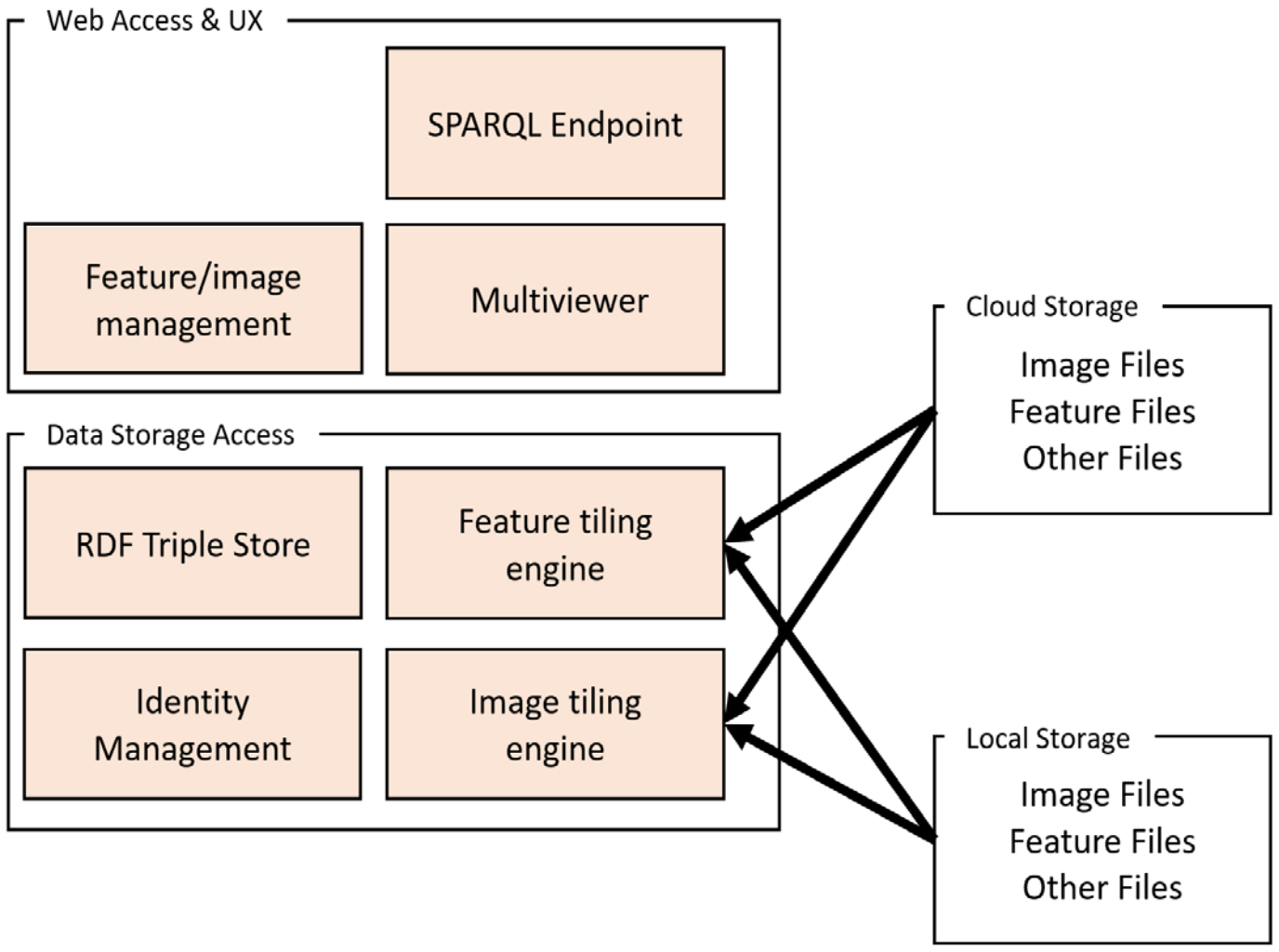

***W3C Open Annotations for images***

The bulk of our annotations are generated using various deep learning whole slide image (WSI) feature extraction pipelines. Different DL frameworks and neural architectures are used and there currently isn't an emergent way to standardize the output of these pipelines with formats used such as Aperio XML, ASAP XML, GeoJSON{Force, 2016 #27}, and PAIS XML{Wang, 2011 #26}.Halcyon adopted earlier in its development the W3C Web Annotations Ontology (https://www.w3.org/TR/annotation-model/) to describe polygonal data which became a W3C recommendation in 2017. The Annotations Ontology is a graph-based data model known as Resource Description Format (RDF) which uses URIs for identifiers. Figure 7 shows a schematic representation of what this data would look like:

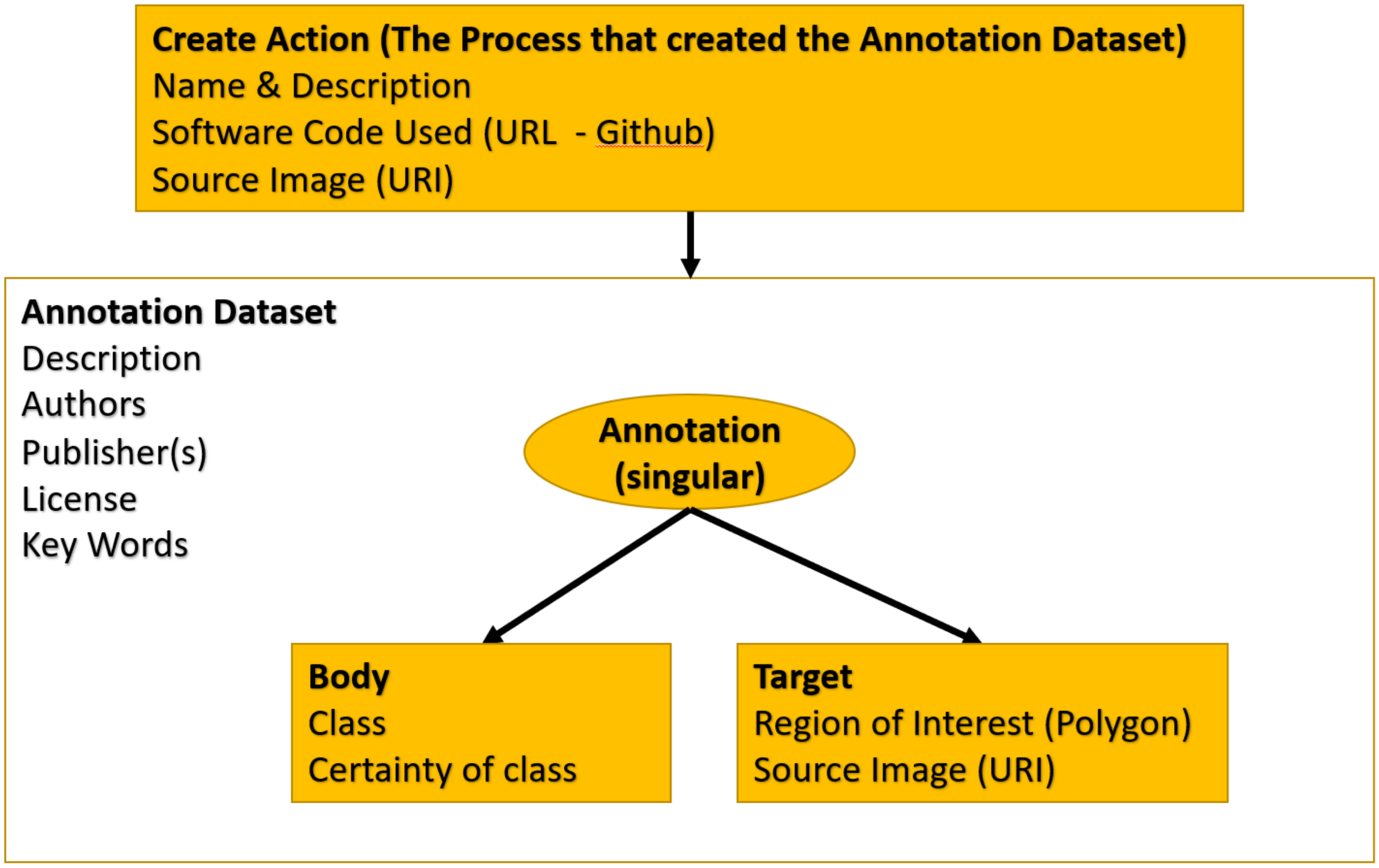

**Figure 7 - W3 Annotations Model contained by RO-Crate model**

We have created a single Extract Transform Load ETL engine to translate that data into the RO-Crate Feature files that will be readable by the Halcyon system. We use SNOMED (https://www.snomed.org/) which provides URIs for their concepts to describe various feature elements like tumor infiltrating lymphocytes, nuclear material (and types), etc. See appendix A for a representation of a singular polygon in W3C Web Annotations in RDF Turtle.

The web annotation data model allows for multiple selectors for an annotation. If we simply want to mark all lymphocytes in a WSI, we can simply repeat the oa:selector and change the polygon values accordingly. The above RDF can be generated in any number of languages including popular languages for deep learning like Python [{, #19}], Javascript [{, #21}], and Java [{, #20}].

The DICOM group has taken up the effort to standardize annotations with DICOM WG26 Supplement 222: Whole Slide Microscopy Annotations{DICOM Standards Committee, 2021 #28}. It's the intent of this project to support this standard with W3C Open Annotations being convertible to Supplement 222. DICOM support is desirable.

***WSI Image Access***

Halcyon has a custom WSI image tiling engine based on ImageBox2{Bremer, 2020 #1}. The primary interface for this image retrieval uses the International Image Interoperability Framework (IIIF) protocol. This protocol was extended here and used for access to the Hilbert curve polygon data. Unifying the two subsystems provides a ready interface for the web-based multi-viewer that is based on OpenSeadragon to use to retrieve data. IIIF's protocol is fairly simple and straight-forward to use, as illustrated by this basic scheme with an example:

{scheme}://{server}{/prefix}/{identifier}/{region}/{size}/{rotation}/{quality}.{format}

Example:
http://server.com/iiif/image.svs/25000,25000,10000,10000/512,512/0/default.jpg

In this example, a JPEG image 512x512 scaled version of a 10000x10000 pixel area with an upper corner at 20000,20000. Now in our new feature engine, we would use something similar:

http://server.com/halcyon/image.svs/25000,25000,10000,10000/512,512/0/default.png

This would return a grayscale image with an alpha channel containing an image of any polygons in this region. If we wanted to access an alternate representation of the polygon data, we could use:

http://server.com/halcyon/image.svs/25000,25000,10000,10000/512,512/0/default.json

In this example, the same polygon data would be returned but in JSON-LD.

Like Whole Slide Image pyramids, Halcyon implements a "feature pyramid" with consecutively reduced polygon maps (Figure 8)). IIIF calls Halcyon to trigger a SPARQL query to retrieve the pertinent polygons at the relevant scale. The data is etched onto a transparent 4-channel PNG making the the primary delivery method to the web viewer being the images. The PNG alpha channel is used to indicate the null case. The Red channel is used to either represent a particular feature class and the Green channel used to indicate probability of a particular class. [0-255] values of gray scale represent probabilities in 0.392% increments which is sufficient for visualizations. If original polygon data is required by the browser/client, change the default image type in the IIIF call from default.png to default.json and the server will return json-formatted polygon data.

Viewers like OpenSeadragon use this image pyramid to only grab images at a sufficiently high resolution to minimize how much data is transferred from the image tiling engine. This approach is here extended to Hilbert features, where the original polygon space is now also scaled by 2, while retaining the efficiency of reducing the polygons retrieved by dropping polygons that are not visible at each scaling. As a consequence, when the viewer asks for polygons at lower resolutions than those already available, the amount of polygons retrieved is reduced and speeds up the baking of the polygons onto the transparent image generation.

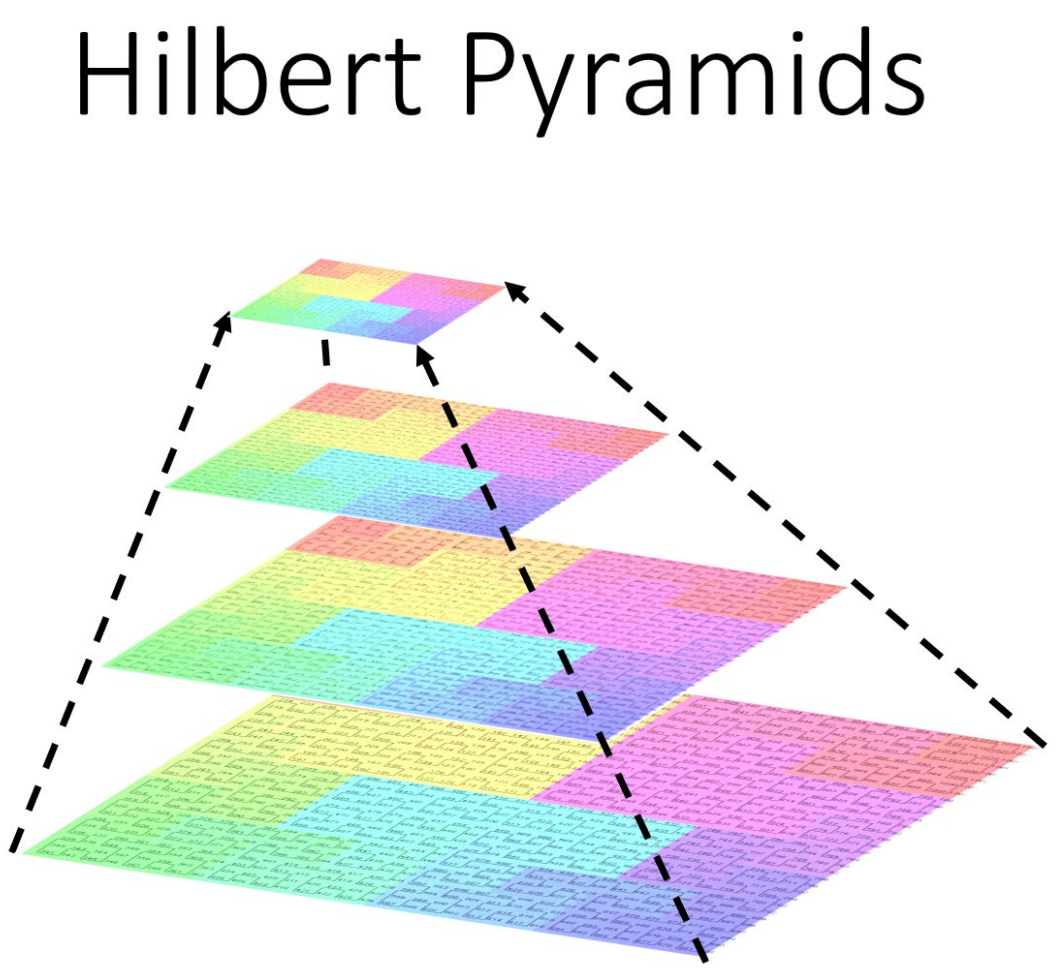

**Figure 8 - Hilbert Pyramid**

Halcyon can store polygons and annotations as RDF that can then be consumed by projects such as BeakGraph,which implements an Apache Jena (https://jena.apache.org) graph backed by Apache Arrow (https://arrow.apache.org). The Apache Arrow data and corresponding index files for a particular RDF graph are grouped and stored in a Research Object Crate (RO-Crate) {Soiland-Reyes, 2022 #37} formatted ZIP file. Implementing the Arrow files as an Apache Jena graph allows us to take advantage of Apache Jena's ARQ engine and allows SPARQL to be used to query our Apache Arrow files. This was done similarly by the RDF HDT (https://www.rdfhdt.org/) project with a different binary storage mechanism. Language bindings for Apache Arrow files are available in multiple languages such as Java, Python, C/C++, Rust, and Javascript. Our Apache Arrow-backed Jena graph implementation (BeakGraph) is only available in Java (https://github.com/ebremer/BeakGraph). However, the data files are still readable with Apache Arrow's other language bindings with only the ability to use SPARQL being lost.

**Data Validation**

RDF allows for data to be serialized in a variety of different ways, whether it be the property or relationship or whether the data itself is represented as a string, float, or integer. Here we will use the W3C Shapes Constraint Language (SHACL) https://www.w3.org/TR/shacl/ to validate data inbound to be loaded into any particular deployment of Halcyon.

***Data Security and Access Control***

Halcyon was designed as a multi-user system with group collaborations. Following the core design choice of W3C RDF linked data, we use the W3C Web Access Control (WAC) vocabulary to define our security Access Control Lists (ACL). When the Halcyon engine processes any and all requests, the access credentials of the requesting agent are checked against the WAC data to validate whether that agent is allowed to read and/or modify the data. Halcyon uses an embedded Keycloak server (https://www.keycloak.org/) for defining users, groups, and roles. The primary authentication method is OpenID Connect. We have extended the keycloak schema for users to include a WebID (https://www.w3.org/wiki/WebID) which is a URI which is used instead of an email address or username string. The WebID provides an identifier for user data and for the WAC ACL data. It is also a precursor for the implementation of WebID-OIDC (https://github.com/solid/webid-oidc-spec) within the Halcyon framework.

**Multi-Viewer**
An essential part of Halcyon, is Multi-Viewer, an interactive, web-based tool for pathologists to help them view and annotate whole slide images. Multi-Viewer is capable of a single or MxN viewports enabling the user to select how they want to view the image(s) and related features layers. From there, the user can interact with it and toggle feature layers on and off, or change the transparency as they wish.

For example, let's say our users want to get four different perspectives of the same image on the screen. Halcyon will invoke the multi-viewer program and request: How many viewers to be displayed, the size of each viewer, the whole slide image with features to be overlayed, the colors and value ranges to be used in representing each feature overlay, and the starting opacity of those features. *(See image below.)*

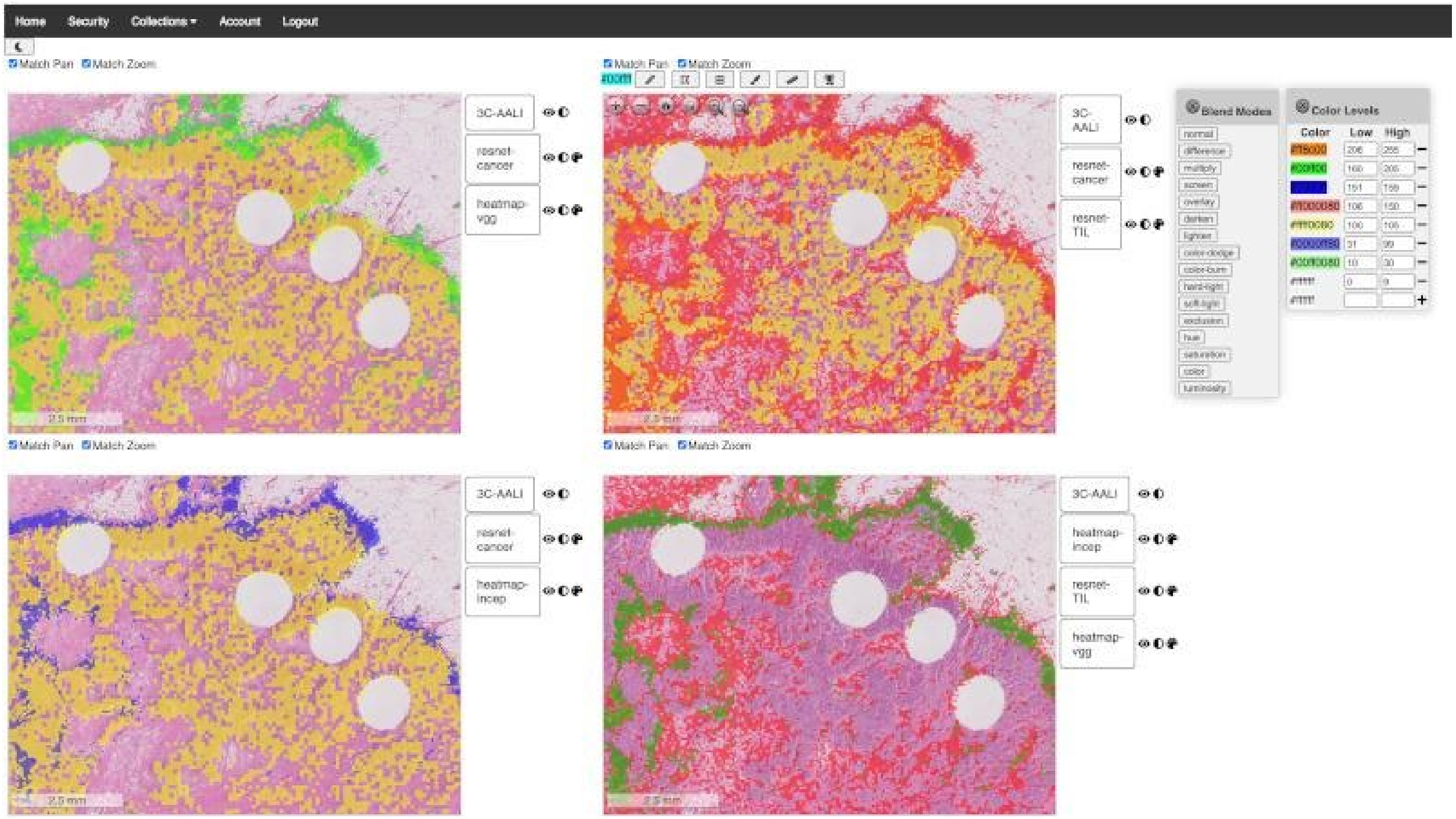

**Figure 9 - 2x2 layout with superimposed feature layers**

All of the viewers are synchronized to pan and zoom to the same location and resolution. Any viewer can automatically become the controller of the other three. Then, the pathologist may decide to draw a polygon, or to measure something of interest, etc.

When viewing the feature layers, they may decide to add, change, or remove the colors and the ranges that they apply to. As the user makes those adjustments, they can see the result of those changes in real time. They can adjust the transparency to their liking in order to see what lies underneath any of the semi-transparent feature overlays.

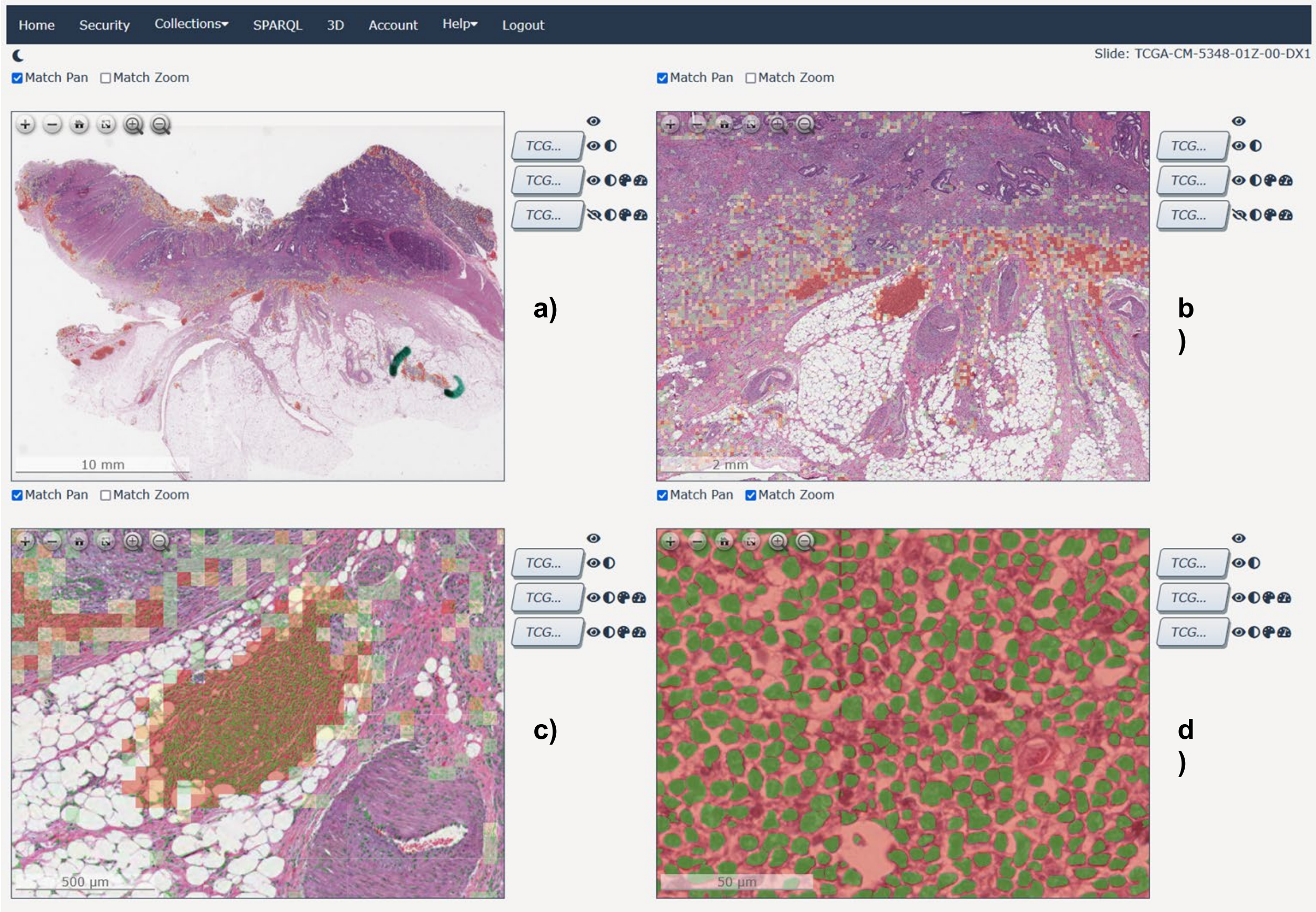

Figure 10 - Nuclear segmentation and Tumor Infiltrating Lypnocytes (TILs) result sets at four different scales. a) Full WSI with heatmap of TILs b) Closer view with panning sync'd c) Even closer view, but now with nuclear segmentation results superimposed d) maximum zoom.

## Discussion and Future work

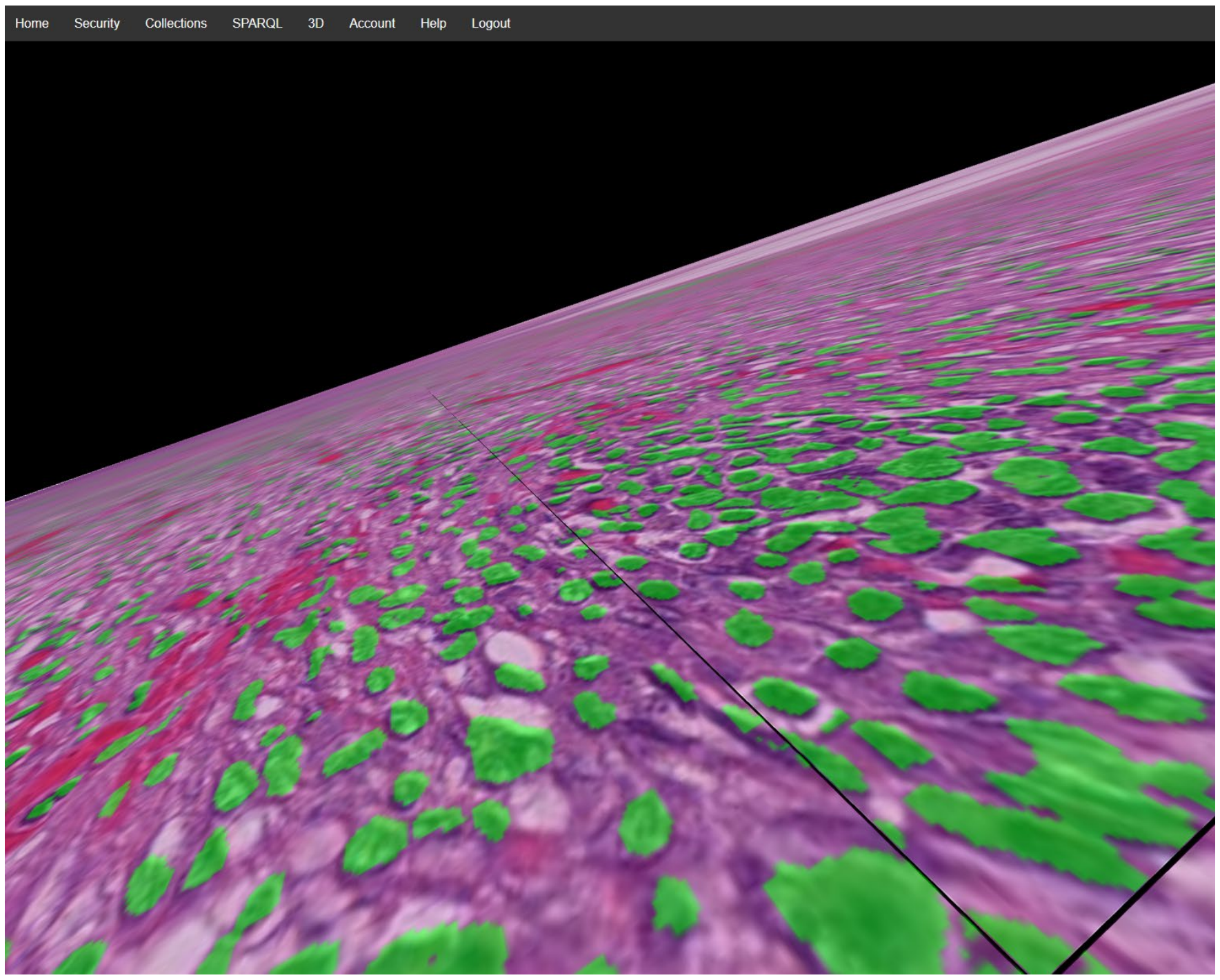

**Figure 11 - WSI in rotated position to display variable resolution using Zephyr prototype**

The scale independent indexing of hilbert curves, and the management of the polygon data using W3C's RDF, offers a number of opportunities for further development:

1) WebGL-based 3D viewer (Zephyr, Figure 11) to support registered stacks of Whole Slide Images and features.  Zephyr uses a variable resolution display providing maximum WSI resolution near the user's point of view (POV)  and lower pyramid resolutions for WSI tiles further from the user POV minimizing the load on the image tiling server.
2) Native support for DICOM dual-personality (DP) image files{Clunie, 2019 #36}.  Although the format emulates both TIFF and DICOM standards, the presence of an embedded image pyramid in the main file is not guaranteed. Image pyramids may be present but are held in separate DICOM files.  This presents a problem for "legacy" viewers which usually expect an image pyramid and that it is present in a singular file.  There is motivation to read the DP files as native DICOM, not only to connect and associate the multi-file image pyramid, but to take advantage of the rich metadata present in DICOM files.
3) extension of HTTP range requests to RO-Crate feature files to remove file locality dependency and enable cloud storage URIs for both Images and Features. {Bremer, 2020 #1} {Bhawsar, 2022 #38}
4) Remote SPARQL-Endpoint support

**Conclusions**

In this report, we describe the software design of our new whole slide imaging and feature management system.  This system leverages Hilbert Curve based data management systems to support efficient storage, query, retrieval and visualization of multi-resolution polygon based bulk annotation datasets. End users can use Halcyon locally or as a multi-user web application for data sharing.  Multiple viewer capability allows for multiple features to be displayed simultaneously whether overlaid on top of each other or side-by-side.  Overlays can be color coded depending on class or probability.  Database interface allows users to query all accessible data and download in multiple formats such as CSV.  An embedded Keycloak server provides authentication and identity management.  Conceptual contributions by Halcyon are as follows:

1) Polygonal and Spatial representation using Hilbert Curves.
2) Adoption of the W3C Annotation ontology for marking up regions of interest
3) Integration of ImageBox2 tiling engine for cloud-friendly image access using HTTP Range requests.
4) Adoption of Linked Data Architecture.

**Code Availability**
Current source code and documentation for Halcyon are available at https://github.com/halcyon-project.

**Deployment**
Halcyon is developed in Java using the JDK17 LTS allowing it to be run cross-platforms.  Installation artifacts for each major OS platform (Windows, Linux, MacOSX) are being generated.  Halcyon has specifically been developed with the GraalVM project (https://www.graalvm.org/) which allows various Halcyon components to be compiled into a native executable for each major OS platform.  No dependencies on additional libraries, nor a docker sub-system required.

**Financial Support and sponsorship**
Partial Support from NCI award 5U24CA215109

**Competing Interests**
There are no conflicts of interest

**Authors' Contributions**
Erich Bremer - Conceptualization, system design, architecture, programming, writing and editing
Tammy DiPrima - Multi-viewer design and coding, writing and editing
Joseph Balsamo - Access control and identity server communications programming
Jonas Almeida - Validation with use cases, writing and editing
Rajarsi Gupta -  Provided use cases, suggested experiments, writing and editing
Joel Saltz -  Provided use cases, suggested experiments, writing and editing

**Appendix A**

```
@prefix exif: <http://www.w3.org/2003/12/exif/ns#> .
@prefix hal:  <https://www.ebremer.com/halcyon/ns/> .
@prefix oa:   <http://www.w3.org/ns/oa#> .
@prefix sno:  <http://snomed.info/id/> .
@prefix so:   <https://schema.org/> .
@prefix xsd:  <http://www.w3.org/2001/XMLSchema#> .

<>      a so:Dataset ; so:creator <http://orcid.org/0000-0003-0223-1059> ;
        so:datePublished "2022-09-16 16:05:56" ;
        so:description "Nuclear segmentation of TCGA cancer types" ;
        so:keywords "pathology" , "Whole Slide Imaging" , "nuclear segmentation" ;
        so:license <https://creativecommons.org/licenses/by-nc-sa/3.0/au/> ;
        so:name "cnn-nuclear-segmentations-2019" ;
        so:publisher <https://ror.org/05qghxh33> , <https://ror.org/01882y777> .

<https://bmi.stonybrookmedicine.edu/nuclearsegmentation/tcga/2019> a so:CreateAction ;
        so:description "cnn-nuclear-segmentations-2019" ;
        so:instrument <https://github.com/SBU-BMI/quip_cnn_segmentation/releases/tag/v1.1> ;
        so:name "cnn-nuclear-segmentations-2019" ;
        so:object <urn:md5:7fdc7b803b9649d2c63d4b5f71419f8f> ;
        so:result <> .

<urn:md5:7fdc7b803b9649d2c63d4b5f71419f8f> exif:height  "78080"^^xsd:int; exif:width
"37120"^^xsd:int .

[ a oa:Annotation ; oa:hasBody [
        a hal:ProbabilityBody ; hal:assertedClass  sno:261665006 ; hal:hasCertainty "1.0"^^xsd:float ] ;
        oa:hasSelector  [ a oa:FragmentSelector ;
                          dcterms:conformsTo <https://www.ogc.org/standard/wkt-crs/> ;
                           oa:hasSource  <urn:md5:7fdc7b803b9649d2c63d4b5f71419f8f> ;
                           rdfs:value "POLYGON ((22284 70778, 22281 70781, 22281 70783, 22283
70785, 22284 70785, 22286 70787, 22288 70787, 22289 70786, 22291 70786, 22292 70785, 22293
70785, 22293 70783, 22291 70781, 22291 70780, 22290 70780, 22289 70779, 22286 70779))" ]] .
```